\documentclass[10pt, a4paper,twocolumn]{article}
\usepackage{xurl}
\usepackage[square,numbers]{natbib}
\usepackage{abstract}
\usepackage{lipsum}
\usepackage{amsmath}
\usepackage{graphicx}
\usepackage{tabularx}
\usepackage{soul}
\usepackage{geometry}
\usepackage[normalem]{ulem}
\usepackage[dvipsnames]{xcolor}
\usepackage{authblk}
\usepackage{array}
\usepackage{mathtools}

\title{Holographic QCD Running Coupling Constant from the Ricci Flow}
\author[a]{H\'ector Cancio}
\author[a, b]{Pere Masjuan}
\affil[a]{\textit{\small{Grup de F\'isica Te\`orica, Departament de F\'isica, Universitat Aut\`onoma de Barcelona and Institut de F\'isica d'Altes Energies (IFAE), 08193 Bellaterra (Barcelona), Spain}}}
\affil[b]{\textit{\small{ Institut de F\'isica d’Altes Energies (IFAE) and The Barcelona Institute of Science and Technology (BIST), Campus UAB, 08193 Bellaterra (Barcelona), Spain}}}

\date{}

\usepackage{hyperref}
\hypersetup{colorlinks=true, urlcolor=blue, linkcolor=blue, citecolor=blue}

\begin{document}
\twocolumn[
\begin{@twocolumnfalse}
\maketitle
\vspace{-3em} 
\begin{abstract}
Through a holographic model of QCD, we present a phenomenological approach to study the running of the strong coupling constant $\alpha_s$ in both non-perturbative and perturbative regimes. The renormalization of the metric tensor, driven by the Ricci Flow, and the breaking of conformal and chiral symmetries -thanks to introducing a double dilaton model and large-$N_c$ corrections- allow us to relate the existence of an infrared fixed point in the coupling constant with a smooth matching to pQCD well above 2 GeV. This is done through a model with two fit parameters and one matching point. The proposed dilaton model yields linear Regge trajectories and decay constants for scalar, vector, and tensor meson families similar to their experimental counterparts.
\end{abstract}
\vspace{2em}
\end{@twocolumnfalse}
]
\vspace{3em}
\section{Introduction}
AdS/CFT conjecture has been used extensively in the search for a possible holographic model dual to QCD in the non-perturbative regime. First introduced as an equivalence between strongly coupled $\mathcal{N}=4$ super Yang-Mills theory and supergravity \cite{Maldacena,Gubser,Witten}, it was shown later it was possible to break conformal symmetry to obtain approximate large-$N_c$ phenomenological models of QCD at low energies. In particular Hard \cite{Erlich,DaRold} and Soft Wall \cite{Karch} models provide a way to compute masses, decay constants, and Regge trajectories of mesons of arbitrary spin. In both cases, their phenomenological scope is still limited \cite{Leutgeb}, yet some advances are promising \cite{Solomko}.

In this work, we are interested in obtaining the running coupling constant in the strongly coupled regime and extending it to the well-known perturbative one. We can get a QCD-like running from dual models like Einstein-dilaton gravity \cite{Gursoy,Kiritsis}, or by considering Light-Front QCD embedded in a AdS background \cite{Brodsky}. In our case we will explore the QCD running coupling constant from a different procedure: we will relate it to the running of the metric tensor on the other side of the duality. Intuitively, an abrupt change of the running coupling constant at $\Lambda_{QCD}$ could be related to an expansion of spacetime on the holographic side, which can be thought of as an exponential ansatz for the running of the metric tensor driven by the so-called Ricci Flow (\cite{Friedan}, \cite{Hamilton} and \cite{Perelman}). The procedure above provides for a family of dilaton fields which beyond returning a running strong-coupling constant will, as a byproduct, return masses and decay constants of vector mesons. When comparing to experimental data points, our objective is then to obtain a well-behaved strong coupling constant at all energy scales containing IR and UV regimes information, rather than predict them.

This letter is organized as follows. In Section \ref{Sec.RicciFlow}, we present some generalities of the Ricci Flow and its relation to the QCD running coupling constant. In Section \ref{Sec.DDSW}, we propose a new holographic QCD model, and we explore its running coupling constant. A discussion of its prediction for masses and decay constants is presented in Section \ref{Sec.MassesDecays}, including a comparison of previous results, and we finalize the letter with concluding remarks in Section \ref{Sec.Conclusions}.

\section{Ricci Flow}\label{Sec.RicciFlow}

The objective is to establish a precise novel relation between the running of the metric tensor and the strong coupling constant $\alpha_s$. In the process, we derive the well-known formula for the Ricci Flow, and we do it so that this derivation ultimately implies the relation as mentioned above between $g_{\mu\nu}$ and QCD $\beta$-function.

Consider a generic AdS background of the form:
\begin{equation}
\label{eqn:AdSBackground}
ds^2=\frac{R^2}{z^2}\left(dz^2+\eta_{\mu\nu}dx^{\mu}dx^{\nu}\right).
\end{equation}
Here $R$ is the AdS radius, which is related to 't Hooft's coupling $\lambda$ by the formula:
\begin{equation}
\label{eqn:Maldacena}
\frac{R^2}{\alpha'}=\sqrt{\lambda},
\end{equation}
where $\lambda=g_{\text{YM}}^2N_c$. We are considering the large-$N_c$ limit in the quantum field theory at the boundary of AdS, so we assume $\lambda \gg 1$. Our interest is in QCD so $g_{\text{YM}}$ is the QCD strong coupling. In this way, since $\alpha_s=g_{\text{YM}}^2/(4\pi)$, from Eq.(\ref{eqn:Maldacena}) we can write:
\begin{equation}
\label{eqn:AdSCFTdictionary}
\alpha_s=\frac{R^4}{4\pi N_c \alpha'^2}.
\end{equation}
If the boundary theory has exact conformal symmetry then $\alpha_s$ is constant and the background remains a pure AdS spacetime. If we want to break conformal symmetry, we can consider a deviation from pure AdS imposing $R=R(z)$, thus considering Eq.(\ref{eqn:Maldacena}) holds approximately. In this way, $\alpha_s$ will depend on the holographic coordinate. More precisely, if we return to Eq.(\ref{eqn:AdSBackground}) and write it in the form
\begin{equation}
\label{metric}
ds^2=\frac{R^2}{z^2}dz^2+g_{\mu\nu}dx^{\mu}dx^{\nu},
\end{equation}
then $g_{\mu\nu}=(R^2/z^2)\eta_{\mu\nu}$ where $R^2\sim\sqrt{\alpha_s}$, see Eq.(\ref{eqn:AdSCFTdictionary}). We can now relate a running of $\alpha_s$ with a running of $g_{\mu\nu}$. In terms of $\mu= z^{-1}$, the renormalization group equation for the metric tensor $g_{\mu\nu}(\mu)=\mu^2R^2(\mu)\eta_{\mu\nu}$, with $R^2(\mu)= \alpha'\sqrt{4\pi N_c\alpha_s(\mu^2)}$, will be:
\begin{equation}\label{eq:RGE}
\mu \frac{\partial g_{\mu\nu}}{\partial \mu}=\Biggl( 2\mu^2 R^2(\mu)+2\mu^3 R(\mu)\frac{d R(\mu)}{d\mu}\Biggr)\eta_{\mu\nu}.
\end{equation}
Multiplying and dividing by $R(\mu)$ in the second term we can write:
\begin{equation}\label{eq:RGE2}
\begin{split}
\mu \frac{\partial g_{\mu\nu}}{\partial \mu}&=\Bigl(2g_{\mu\nu}(\mu)+2\mu g_{\mu\nu}(\mu)\frac{d}{d\mu}\log R(\mu)\Bigr)\\
&=2g_{\mu\nu}(\mu)\Bigl(1+\frac{\mu}{4}\frac{d}{d\mu}\log (4\pi N_c\alpha'^2\alpha_s(\mu^2))\Bigr)\\
&=2g_{\mu\nu}(\mu)\Bigl(1+\frac{\mu}{4}\frac{d}{d\mu}\log (\alpha_s(\mu^2))\Bigr).
\end{split}
\end{equation}
In this way, we can identify the second term $\beta_{\mu\nu}=g_{\mu\nu}\frac{\mu}{4}\frac{d}{d\mu}\log \alpha_s$ as an homologous to a $\beta$-function of the metric $g_{\mu\nu}$, and relate naturally the running of $\alpha_s(\mu^2)$ with the running of the metric tensor $g_{\mu\nu}(\mu)$. This second running is, at first order in $\alpha'$, $\beta_{\mu\nu}=-2R_{\mu\nu}$, where $R_{\mu\nu}$ is the Ricci tensor of the metric $g_{\mu\nu}$ \cite{Friedan}. This result is specific for the metric Eq.(\ref{metric}), but motivates the introduction of the running of a general metric tensor as:
\begin{equation}
\frac{\partial g_{\mu\nu}}{\partial t}=2g_{\mu\nu}-2R_{\mu\nu},
\end{equation}
where $t=\log{\mu}$. This formula is known as the Ricci Flow. Introduced in the context of string theory \cite{Friedan}, and in mathematics \cite{Hamilton,Perelman}, Ricci Flow has been a method to compute the renormalization of non-linear sigma models via the above equation. Also, we can conceive the running of the metric tensor as the radial evolution equation in Hamilton-Jacobi formulation (see \cite{Jackson} for more details about this connection). In this direction we will treat the holographic coordinate as the evolution parameter of the flow.

From the previous results, if $\alpha_s$ does not run at all, solving the Ricci Flow should return a metric tensor corresponding to AdS background: assume $R_{\mu\nu}=0$, then the solution for the flow equation
\begin{equation}
\frac{\partial g_{\mu\nu}}{\partial t}=2g_{\mu\nu}
\end{equation}
is, in terms of the holographic coordinate $z$,
\begin{equation}
g_{\mu\nu}(z)=\frac{1}{z^2}\eta_{\mu\nu},
\end{equation}
considering $R=1$. We obtain the expected AdS background:
\begin{equation}
ds^2=\frac{1}{z^2}\left(dz^2+\eta_{\mu\nu}dx^{\mu}dx^{\nu}\right). 
\end{equation}
For a non-trivial running of $\alpha_s$, we have to break conformal symmetry in the geometric side of the duality by performing a deformation from the AdS metric. Consider the parametrization $R_{\mu\nu}=\Lambda(t)g_{\mu\nu}$ for $\Lambda(t)$ an unknown function. In this way we must solve the flow:
\begin{equation}\label{flow}
\frac{\partial g_{\mu\nu}}{\partial t}=2\left(1-\Lambda(t)\right)g_{\mu\nu}.
\end{equation}
As observed in the introduction, an abrupt change of $\alpha_s$ at low energies could be related to an expansion in AdS spacetime, cf. Eq.(\ref{eqn:AdSCFTdictionary}), since growth in $\alpha_s$ implies an increase in AdS radius $R$. The function $\Lambda(t)$ will ultimately determine the warp factor of the metric tensor of the model and the running of $\alpha_s$. From all the possibilities we have chosen the one that induces the fastest growth yet keeping a simple ansatz, to obtain an abrupt change of $\alpha_s$ at low energies. Therefore, for simplicity, we assume an exponential varying function of the form $\Lambda(t)=c_0 e^{c_1 t}$, where $c_0$ has units of energy squared and $c_1$ is unitless. Other possibilities such as a polynomial ansatz, would induce a slower growth. We softly break the conformal symmetry by allowing $t$ dependence in $\Lambda$. The exponential parameterization will enable us to perform the desired abrupt change of $\alpha_s$ at low energies. Other models of conformal symmetry breaking are possible as, for example, we could model $\Lambda(t)$ as a step function to represent Hard Wall models.

We stick to Eq.(\ref{flow}) as it can be considered a convenient ansatz for solving the flow. Using the aforementioned exponential dependence, Eq.(\ref{flow}), leads us to:
\begin{equation}
g_{\mu\nu}(t)=\eta_{\mu\nu}\text{exp}\left(2 t -2\frac{c_0}{c_1}e^{c_1 t}+k\right),
\end{equation}
where $k$ is an integration constant. In conformal coordinates ($t=-\log z$):
\begin{equation}
g_{\mu\nu}(z)=\eta_{\mu\nu}\text{exp}\left(-2\log{z} -2\frac{c_0}{c_1}z^{-c_1}+k\right).
\end{equation}
To fix the constants $c_0$ and $c_1$, we notice that the volume form present in the Soft Wall action is typically $\exp(-z^2)/z^5$ \cite{Karch}. Therefore, we fix $c_0=2/5 \text{ GeV}^{2}$ to reproduce that usual volume form, and $c_1=-2$ to obtain the correct $z^{-2}$ term. We obtain:
\begin{equation}
g_{\mu\nu}(z)=\frac{e^{c_0 z^2+k}}{z^2}\eta_{\mu\nu}.
\end{equation}
This kind of background improving the original Soft Wall model have been used recently in \cite{MartinContreras}, \cite{Capossoli}, \cite{Rinaldi} and \cite{Ceccopieri} for different kind of purposes. In our case, we require it in order to study the strong coupling constant $\alpha_s$. The interpretation is clear: $\Lambda=\Lambda(t)$ leads to the presence of a dilaton $\phi(z)$, the field in charge of conformal symmetry breaking in the holographic model \cite{Karch} as deformation of the metric tensor. An exponential ansatz is enough to find a Soft Wall-like dilaton plus a constant. The integration constant $k$ could be interpreted as an AdS radius $R_0^2$, giving a warp factor of the form $R_0^2 e^{-(2/5)\phi(z)}/z^2$ if $k=2\log{R_0}$, but for now, let us consider it part of the dilaton $\phi$.

With the previous choice of $c_0$ and $c_1$, we obtain the following volume form:
\begin{equation}
\sqrt{g}=\frac{e^{-\phi(z)}}{z^5}.
\end{equation}
The denominator gives the measure, and the numerator is the usual dilaton term in the action of holographic models \cite{Karch}.

As a consistency check, we observe that if $\phi(z)=\phi(t(z))$ is the obtained dilaton, then $d\phi(t)/dt=-2\Lambda(t)$. In this way, the function that induces a deviation from pure AdS spacetime, $\Lambda(t)$, is related directly to the field that breaks conformal symmetry in the holographic model, $\phi(z)$.

As summary, in presence of a dilaton $\phi(z)$, we identify the AdS radius of Eq.(\ref{eqn:AdSCFTdictionary}) as $R^4(z)=\exp{(-(4/5)\phi(z))}$. As a consequence, we have a $z$-dependent coupling constant:
\begin{equation}
\label{eqn:RunningDilaton}
\alpha_s(z)=\frac{1}{4\pi N_c \alpha'^2}e^{-\frac{4}{5}\phi(z)}.
\end{equation}
From this viewpoint, Hard Wall models are obtained after considering a step function for $\exp{(-\phi(z))}$. 

The above procedure allows us to define a general QCD running coupling constant for models with a dilaton. In other words, each holographic model will induce a particular running of $\alpha_s$ as, for example, the one obtained in Ref.\cite{Brodsky} using an embedding of Light-Front QCD into a Soft Wall model. In the next section, we propose a dilaton model that will induce a QCD running coupling constant with an infrared fixed point and with a matching to pQCD. 

\section{Double Dilaton Soft Wall model and its running}\label{Sec.DDSW}

Our attempt is to explore what sort of dilaton field could return a running coupling constant resembling the one in QCD. A confining dilaton was used in \cite{Brodsky}, which had the opposite sign in comparison with the original dilaton presented in \cite{Karch}. We now introduce a model with two dilatons, one with a positive and one with a negative sign, all together in a single model. This choice of signs will be responsible for breaking the chiral symmetry straight away and will enable us to study properties of the running of $\alpha_s$ at the same time. Physically, this setup should correspond to a system with vector and axial-vector mesons. We see that this system naturally provides a running of the strong coupling constant with an infrared fixed point. A Soft Wall dilaton would have an ambiguity sign for this system: Regge trajectories are the same regardless of the sign of the dilaton. To fix the sign of the dilaton, one can relate to higher spin mesons as done in \cite{Katz}. Alternatively, we consider that vector and axial-vector mesons see different geometries \cite{Hirn}.
In reference \cite{Hirn}, different warp factors for vector and axial-vector mesons were introduced in order to reproduce the well-known QCD OPE of the vector and axial-vector two-point correlation functions.
As such, each type of meson sees dilatons with opposite signs \cite{Nicotri,Zuo,Afonin}.
In particular, Ref.\cite{Afonin} remarks on the importance of having opposite dilaton signs for axial-vector mesons. The dilaton sign choice impacts the IR part of the action. Consequently, it seems reasonable to think that sign choices for vector and axial-vector mesons have an impact both in the IR regime where the running of $\alpha_s$ is not yet known, and in the UV, where the connection of $\alpha_s$ with perturbative QCD running have not been fully accomplished. 
So we consider potentially different warped factors to describe dynamics. In this way, think of a background defined in $AdS \times AdS'$. Ricci tensor factorizes in the product, so we can conceive each piece evolves with the equations:
\begin{equation}
\label{eqn:TwoRiccis}
\frac{\partial g_{\mu\nu}}{\partial t}=2g_{\mu\nu}-2R_{\mu\nu}, \text{ } \frac{\partial g'_{\alpha\beta}}{\partial t'}=2g'_{\alpha\beta}-2R'_{\alpha\beta}.
\end{equation}
The evolution allows us to consider two different dilatons, allowing choosing a dilaton $\phi$ in $AdS$ and its opposite $-\phi$ in $AdS'$. This setup enables us to work with a ten-dimensional model in $AdS\times AdS'$ with action:
\begin{equation}
\label{eqn:action}
S=\int d^5xd^5x'\sqrt{g_{\times}}\left(-\frac{1}{2g_5^2}\left(F_L^2+F_R^2\right)\right).
\end{equation}
$x$ and $x'$ are spacetime coordinates in $AdS$ and $AdS'$, respectively. $g_{\times}$ is the determinant of the metric tensor in the product $AdS\times AdS'$. This is an action of a theory of gauge fields $A_L$ and $A_R$ dual to the QCD currents $J_L$ and $J_R$ in ten dimensions. Also, $g_5$ is a Yang-Mills coupling and $F_L=F_L(x)$ and $F_R=F_R(x')$ are field strength tensors defined in the usual way \cite{Yang}.

By using the exponential ansatz for $\Lambda(t)$ used in the previous section
\begin{equation}
\Lambda(t)=\frac{2}{5}e^{-2 t},
\end{equation}
we can obtain dilatons $\phi(z)=-z^2-k$ and $\phi'(z)=+z'^2+k'$, being $k$ and $k'$ integration constants. From this point of view $AdS$ and $AdS'$ will have the following actions, respectively:
\begin{equation}
S=\int d^5x\frac{e^{-z^2}}{z^5}\left(-\frac{1}{2g_5^2}F_L^2\right),
\end{equation}
\begin{equation}
S'=\int d^5x'\frac{e^{+z'^2}}{z'^5}\left(-\frac{1}{2g_5^2}F_R^2\right).
\end{equation}
\\
To obtain a five-dimensional model we consider the diagonal $\Delta$ of $AdS\times AdS'$, defined as the submanifold $x=x'$. The action on the diagonal is given by the sum of the above two actions evaluated at the diagonal:
\begin{equation}
S^{\Delta}=\int d^5x\frac{e^{-\phi^{\Delta}(z)}}{z^5}\left(-\frac{1}{2g_5^2}\left(F_L^2+F_R^2\right)\right).
\end{equation}
This choice not only gives a five-dimensional holographic model but also breaks chiral symmetry. For instance  $AdS$ contains a $U(N_f)_L$ theory and $AdS'$ contains a $U(N_f)_R$ theory, being $N_f$ the number of flavours. The spacetime product $AdS\times AdS'$ contains a theory with gauge symmetry $U(N_f)_L\times U(N_f)_R$, as can be seen from the action in Eq.(\ref{eqn:action}). By selecting the diagonal one obtains a $U(N_f)_V$ theory, breaking chiral symmetry geometrically. This is the only choice in our setup that will give a reasonable holographic model of QCD. The \textit{diagonal} dilaton obtained is given by:
\begin{equation}
\label{eqn:dilaton}
\phi^{\Delta}(z)=\log\left(2\cosh(\lambda^2(z^2+k))\right)\, ,
\end{equation}
where we have used the identity $\cosh(x)=(1/2)(e^{+x}+e^{-x})$. The constant $\lambda$ is introduced to render the argument of $\cosh$ unitless, we have identified $k=k'$ and we have chosen $k=1 \text{ GeV}^{-2}$ for simplicity.
To verify $\int_0^{\infty}e^{-\phi^{\Delta}(z)}dz<\infty$ we consider a positive dilaton. From now on we will refer to this model as the Double Dilaton Soft Wall model (DDSW).

At this point, we shall comment about what would happen if other signs for the dilaton were chosen. If both signs are positive, we would obtain again a Soft Wall-like dilaton as the one considered in \cite{Brodsky}; and if both signs are negative we would obtain a running coupling constant with no physical significance for QCD. In this way, our choice of opposite signs is the appropriate one if matching to pQCD is pursued.

In Ref.\cite{Brodsky} the running of strong coupling constant at low energies was obtained with the embedding of Light-Front QCD into a Soft Wall model with a positive dilaton background, obtaining the so-called Holographic Light-Front QCD model. The properties of a low-energy running coupling constant and its $\beta$-function were studied since low energies are the natural landscape of application of holographic models. In our case, since no intrinsic scale is used in holographic models of QCD, the distinction between low and high energy is blurred, yet a matching with pQCD should be possible. Our objective is to find a running with an infrared fixed point and low energies and approach to pQCD result at high energies. A possible avenue would follow Ref.\cite{TeramondPaul} and match both regimes using an analytical continuation of the gauge/gravity duality using a Holographic Light-Front QCD model. We will follow a different approach and explore such matching from the large-$N_c$ expansion.

To do that, and at leading order in $1/N_c$, let us use the expression deduced from the AdS/CFT dictionary formula seen in Eq.(\ref{eqn:RunningDilaton}).
As usual we identify $\mu \sim Q$ so we can define $\alpha_s(Q)\sim\alpha_s(1/z)$. Using Eq.(\ref{eqn:dilaton}) in Eq.(\ref{eqn:RunningDilaton}) with $N_c=3$, and recalling  experimental data is given in terms of $\alpha_s(Q)$, the DDSW model predicts:
\begin{equation}
\label{eqn:parametrization}
\frac{\alpha_s(Q)}{\pi}=\frac{a}{\cosh^{4/5}{(b(Q^2+1))}}.
\end{equation}

To reach Eq.(\ref{eqn:parametrization}), we do not only embed Eq.(\ref{eqn:dilaton}) in Eq.(\ref{eqn:RunningDilaton}) but also redefine the constants that normalize the function and render the $\cosh$ function unitless. As such, the new parameter $b$ has dimensions $[\text{energy}^{-2}$], and $\lambda$ has dimensions $[\text{energy}$]. Being the constant introduced in Eq.(\ref{eqn:dilaton}), $\lambda$ is analogous to the energy scale one introduces as a constant in the quadratic dilaton that defines the usual SW model. In this way, $\lambda$ should determine the mass spectrum of the model, while $b$ remains as a fit parameter in order to preserve the correct units in the cosh function.

\begin{figure}
\includegraphics[scale=0.55]{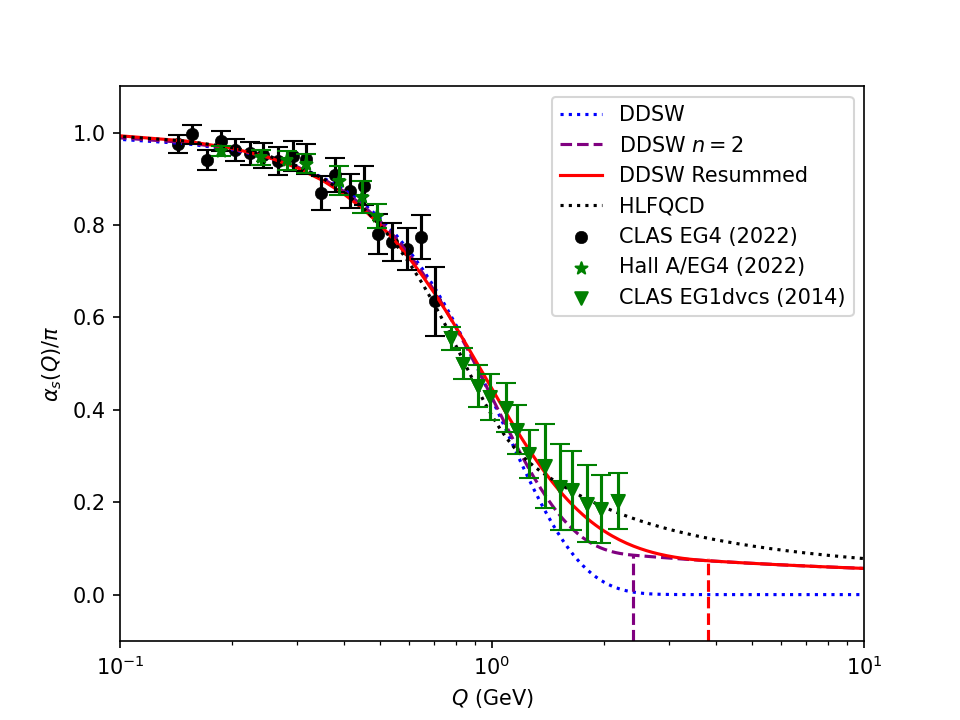}
\caption{Model from Eq.(\ref{eqn:parametrization}) (dotted blue), matching to perturbative QCD running coupling using DDSW model Eq.(\ref{eqn:largeN}) for $n=2$ (dashed purple), resummed (red) Eq.(\ref{eqn:Resum}). Their $\chi^2/DOF$ values are $1.45$, $0.99$ and $0.93$, respectively. We also compare with the holographic light-front model \cite{Brodsky} (dotted black). The dashed vertical lines indicate the matching point with perturbative QCD for $n=2$ and the resummed version (so for $Q>2.39$ and $Q>3.79$, respectively, we have perturbative QCD running coupling in the same color).}
\label{figure:running}
\end{figure}

The parameters $a$ and $b$ are determined after data fitting. We use data for the so-called effective strong coupling or effective charge $\alpha_{g1}(Q)/\pi$ collected in Ref.\cite{Deur}. The effective strong coupling
constant is related to $\alpha_s$ from the perturbative series of an observable truncated to its first order in $\alpha_s$ \cite{Grunberg:1980ja}. Using first-order pQCD equations makes the effective coupling renormalization-scheme and gauge independent, and free of divergence at low $Q^2$. Data in Ref.\cite{Deur} are obtained from JLab experiments Hall CLAS EG4 (from Q = 0.143 GeV to 0.704 GeV), CLAS EG4/E977110 (from Q = 0.187 GeV to 0.490 GeV) and EG1dvcs (from $Q=0.775 \text{ GeV}$ to $2.177 \text{ GeV}$), represented in Fig.\ref{figure:running} as solid star, solid circle, and solid triangle, respectively. Notice the normalization at the fixed-point $\alpha_s(Q \simeq 0)/\pi \simeq 1$ in Fig.\ref{figure:running}. As stated, these data extract the strong coupling constant in the effective charge approach, offering a non-perturbative renormalization-scale invariant treatment of $\alpha_s$ at low energies. From a fit to data with the model in Eq.(\ref{eqn:parametrization}), we obtain $a = 1.545 \pm 0.047$ $b=1.150 \pm  0.041 \text{ GeV}^{-2}$ with a reasonable $\chi^2/DOF = 1.45$. As can be seen in Fig.\ref{figure:running}, we obtain naturally an infrared fixed point. At $Q\simeq 1.5\text{ GeV}$ we observe a slight deviation from the experimental result and a tendency for $\alpha_s(Q)/\pi$ to go to zero quite fast. We have also compared in Fig.\ref{figure:running} with the holographic light-front model (\cite{Brodsky}, $\alpha^{AdS}_{Modified,g1}$ Eq.(12)). The base holographic light-front result is not displayed, being very similar to the DDSW result in blue in Fig.\ref{figure:running}.
 
Even though the fit is reasonably good, we do not find pQCD running coupling constant at high energies. This is expected since holographic models of QCD work only in the strongly coupled regime. At higher energies, the running should have a transition to the well-known perturbative one. Nonetheless, this deviation occurs at higher energies compared to, for example, the Light-Front QCD AdS model employed in \cite{Brodsky}, which found a transition between the AdS and the pQCD running at around 1 GeV. The same authors improve it by joining their model with the pQCD prediction, resulting in a better result in the perturbative regime.

Our running has been deduced from a holographic model based on AdS/CFT duality, so one should expect large-$N_c$ corrections to this result. Since the large-$N_c$ limit corresponds to the non-relativistic limit, corrections to that limit shall allow going beyond the non-relativistic limit. In practice, corrections to the large-$N_c$ limit should allow matching to pQCD to higher scales, beyond 1 GeV. These large-$N_c$ corrections are difficult to derive from first principles. We follow a more phenomenological approach. At finite $N_c$, quark loops are suppressed by $1/N_c$, and non-planar (self-intersecting) diagrams are suppressed by $1/N_c^2$ \cite{'tHooft}. Any observable can be expanded in the form \cite{Zaffaroni}:
\begin{equation}
\sum_{g=0}^{\infty}\sum_{L=0}^{\infty}N_c^{2-2g-L}f_g\left(t^2\right).
\end{equation}
Here $t=\alpha'/R^2$, and $\chi=2-2g-L$ is the Euler characteristic that classifies Feynman diagrams in planar and non-planar diagrams by its embedding on closed surfaces \cite{'tHooft}. $g$ is the genus of the surface and $L$ is the number of quark loops (number of boundaries).
For example, planar diagrams correspond to $g=L=0$ and can be embedded in the sphere ($\chi=2$), while $g>0$ correspond to non-planar diagrams and are embedded in surfaces with holes ($\chi<2$). This classification allows $\chi$ to run from $2$ to $-\infty$, including odd values because of the boundaries. The Ricci Flow is first order in $\alpha'$ so a factor $f_g\left(t^2\right)$ is expected in the expansion.
This expansion can be understood as quantum corrections to the AdS/CFT dictionary on Eq.(\ref{eqn:AdSCFTdictionary})  as \cite{Zaffaroni}:
\begin{equation}
\label{eqn:sumCorrections}
\alpha_s(z)=\frac{R_0^4}{4\pi N_c\alpha'^2}+\frac{R^4(z)}{4\pi N_c\alpha'^2}+\left(\frac{R^4(z)}{4\pi N_c\alpha'^2}\right)^2... ,
\end{equation}
where $R_0$ is the usual AdS radius. We choose the above form to be able to perform a resummation. The expansion parameter we use here is $1/N_c$, considering $N_c\to\infty$ grows faster than $R^2/\alpha'$.
The resummation of pQCD running coupling constant returns a Landau pole at low energies. Since Eq.(\ref{eqn:AdSCFTdictionary}) relates $\alpha_s$ with AdS geometry, a Landau pole in $\alpha_s$ at IR would imply a divergent metric tensor at finite $z>0$ in the UV, which is absent from our formulation. Therefore, the observed finitude of our background metric tensor is fundamental to establishing that $\alpha_s$ at low energies is finite. Imposing a finite metric tensor in the UV returns an IR fixed point in the running of $\alpha_s$. In practice, the first term of the above sum will avoid the Landau pole to find a smooth match with pQCD running. 

Eq.(\ref{eqn:sumCorrections}) can be rewritten using then Eq.(\ref{eqn:RunningDilaton}) as:
\begin{equation}
\alpha_s(z)=\sum_{\chi}e^{-\frac{4}{5}\chi\phi(z)}\left(\frac{1}{4\pi N_c\alpha'^2}\right)^{\chi}.
\end{equation}

\noindent
Now, observe that the term $N_c^{2-b}$ can be resummed and be considered as a normalization constant $c$ for $N_c=3$. With all these considerations, our final parametrization is the following partial sum:
\begin{equation}
\label{eqn:largeN}
\frac{\alpha_s(Q)}{\pi}=c\sum_{g=0}^{n-1} \left(\frac{a}{\cosh^{4/5}{(b(Q^2+1))}}\right)^g.
\end{equation}
Here we have included the term $N_c^{-2g}$ inside $a^g$. Thanks to the $g=0$ term, which allows a horizontal asymptote, we can match to perturbative QCD running at four loops with reference scale $M_Z=91.18 \text{ GeV}$, $\alpha_s(M_z)=0.118$. The parameters $a,b,c$ must be determined either from data or from matching to pQCD or a combination of both. Having three parameters to adjust, we acknowledge the limitations of the model (\ref{eqn:largeN}).  We believe the model could be improved in future works to have more predictive power. In the following, we opt for matching only the parameter $c$ and let the data tell us about $a$ and $b$. By matching, we mean we normalize at $Q_0$ both image and derivative of the non-perturbative and perturbative results for the strong coupling. In this way we determine the $c$ parameter solving a system of equations, while a fit to the data will determine $a$ and $b$ parameters. To determine $c$, we can search for a matching energy point $Q_0$ in which pQCD running coupling, coming from high energies, meets the model of Eq.(\ref{eqn:largeN}), coming from low energies, with the following iterative procedure. However, the matching energy point $Q_0$ is not known, which implies we must solve a system of two equations: 
\begin{equation}
\label{eqn:matchingEq}
\alpha_s^{\text{DDSW}}(Q_0)=\alpha_s^{\text{pQCD}}(Q_0),
\end{equation}
\begin{equation}
\frac{d\alpha_s^{\text{DDSW}}}{dQ}\bigg{|}_{Q=Q_0}=\frac{d\alpha_s^{\text{pQCD}}}{dQ}\bigg{|}_{Q=Q_0}.
\end{equation}
We solve the system recursively, starting with the initial value $c=\frac{\alpha_s(M_z)}{\pi} = \frac{0.118}{\pi} \simeq 0.0376$, for then fit the data to obtain $a$ and $b$. Solving the system we determine $Q_0$. So, if $c$ is unfrozen, the previously determined $Q_0$ would return a new value for $c$ which comes along with a new pair of fit parameters $(a,b)$, and reiterate the process until the distance between two consecutive $c$ parameters is equal to $10^{-10}$. Convergence is very quick, giving concrete values of the $c$ parameter and matching point $Q_0$, see Table \ref{Table:fitresults}.
The first non-constant version of the model in Eq.(\ref{eqn:largeN}) has 3 terms ($n=2$). A fit to data returns $a=18.87 \pm 0.60$, $b=(1.289 \pm 0.041) \text{ GeV}^{-2}$, $c=0.082$ and a $\chi^2/DOF = 0.99$. This result is depicted in Fig.\ref{figure:running} as a purple-dashed line (notice we plot $\alpha_s(Q)/\pi$), with a matching with pQCD at 2.39 GeV. 

\begin{figure}
\includegraphics[scale=0.55]{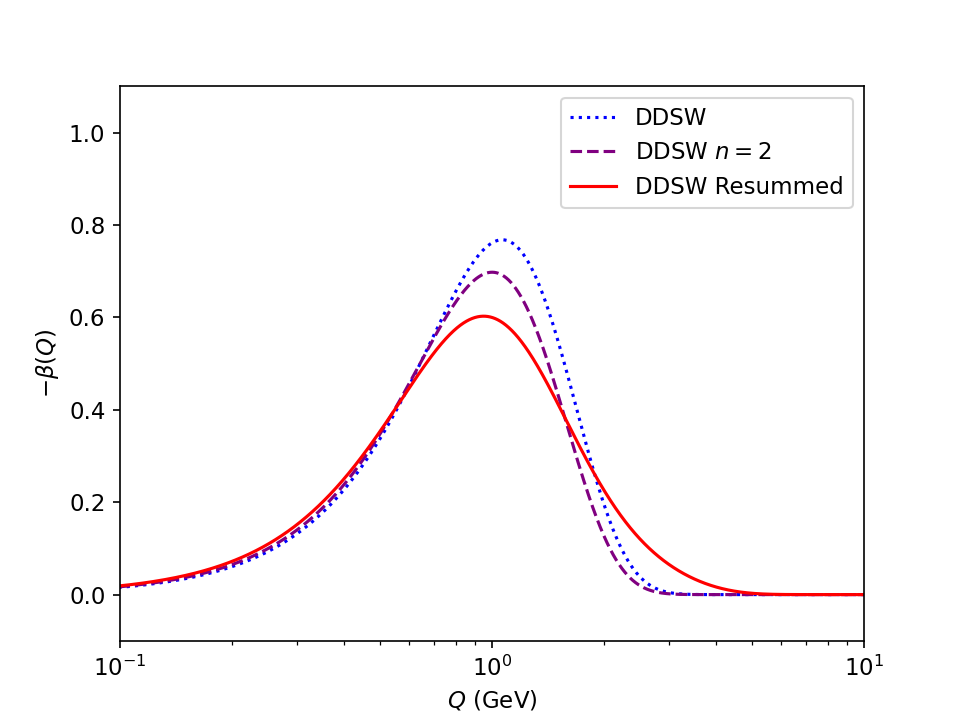}
\caption{$\beta$-function with opposite sign corresponding to Eq.(\ref{eqn:parametrization}) (dotted blue), Eq.(\ref{eqn:BetaDefinitive}) with $n=2$ (dashed purple) and Eq.(\ref{eqn:resummedBeta}) (red).}
\label{figure:beta}
\end{figure}

We can add more and more terms to the partial sum in Eq.(\ref{eqn:largeN}). Since it is a geometric series it can be fully resumed assuming the ratio is less than unity (suggested by the fitted coefficients), to obtain:
\begin{equation}
\label{eqn:Resum}
\frac{\alpha_s(Q)}{\pi}=\frac{c}{1-a \text{ sech}^{4/5}(b(Q^2+1))}.
\end{equation}
In this case, and with the fitted coefficients shown in Table \ref{figure:beta}, the matching point is moved up to $Q=3.79 \text{ GeV}$, see Fig.\ref{figure:running} and Table \ref{Table:fitresults}. Interestingly, the resummation has a slower decrease at around $1-2$ GeV, allowing a better fit to the data.

 In the end, requesting a matching at a given energy point implies the non-perturbative running is predicting perturbative $\beta$-function coefficients. Indeed, using the fitted values for $a$, $b$ with fixed values for $c$ and $Q_0$ in Eq.(\ref{eqn:Resum}) we can compute the $\beta_0$ coefficient using Eq.(\ref{eqn:matchingEq}) with the 1-loop pQCD running coupling constant:
\begin{equation}
\beta_0=\frac{1 - a \text{ sech}^{4/5}(b (Q_0^2 + 1))-\frac{\pi c}{\alpha_s^{\text{pQCD}}(M_Z)}}{c \log\left(\frac{Q_0^2}{M_Z^2}\right)}
\end{equation}
We obtain $\beta_0 = 2.04 \pm 0.01$, to be compared to the theoretical pQCD result $\beta_0=1.917$ for $N_f=5$, so we obtain a departure below $10\%$. The higher-order $\beta$-function coefficients $\beta_1$, $\beta_2$ and $\beta_3$ can be obtained using a Monte Carlo method and the corresponding pQCD expression for $\alpha_s$, but are more difficult to determine since they are suppressed by powers of $\alpha_s(M_z)/\pi$, so they can be predicted with huge errors only.
We could also attempt to predict the leading-order term of the QCD $\beta$-function in the large-$N_c$ expansion. Again, current experimental uncertainties mask the significant results.

We can also determine the non-perturbative $\beta$-function analytically following:
\begin{equation}
\beta(Q)= Q\frac{d}{d Q}\frac{\alpha_s(Q)}{\pi}.
\end{equation}
For a finite $n$ in Eq.(\ref{eqn:largeN}), $\beta$-function reads:
\begin{equation}
\label{eqn:BetaDefinitive}
\begin{split}
\beta(Q)&=-\frac{8}{5}bc Q^2\text{ tanh}(b(Q^2+1))\\
&\cdot\sum_{g=1}^{n-2}g\left(\frac{a}{\text{cosh}^{4/5}(b(Q^2+1))}\right)^{g}.
\end{split}
\end{equation}
\begin{table}[h!]
\begin{center}
\begin{tabular}{ |c|m{1.3cm}| m{1.3cm}|c|m{1.2cm}|c| } 
 \hline
 $n$ & \centering$a$ & \centering$b$ [$\text{GeV}^{-2}$] & $c$ & \centering $Q_0$ [GeV] &$\chi^2$ \\ \hline
 $2$  & \centering$18.8(6)$ & \centering$1.29(4)$ & $0.082$ &  \centering$2.39$ & $0.99$\\
\hline
$3$  & \centering$3.80(5)$ & \centering$0.86(2)$ & $0.080$ & \centering$2.53$ & $0.97$\\
\hline
$4$& \centering$2.22(2)$ & \centering$0.71(1)$ & $0.079$ & \centering$2.63$ & $0.96$\\
\hline
$5$& \centering$1.71(1)$ & \centering$0.62(1)$ & $0.077$ & \centering$2.72$ & $0.95$\\
\hline
$10$& \centering$1.149(4)$ & \centering$0.45(1)$ & $0.074$ & \centering$3.07$ &  $0.90$\\
\hline
$100$ & \centering$0.963(1)$ & \centering$0.297(7)$ & $0.070$ & \centering$3.79$ &$0.92$\\
\hline
$\infty$ & \centering$0.962(1)$ & \centering$0.297(7)$ & $0.070$ & \centering$3.79$ & $0.93$\\
 \hline
\end{tabular}
\end{center}
 \caption{Fit of parameters $a$ and $b$ for a fixed number of terms from $0$ up to $n-1$ in the running of Eq.(\ref{eqn:largeN}) and results for the $c$ parameter and matching point $Q_0$. The last row corresponds to the resummation of Eq.(\ref{eqn:Resum}).}\label{Table:fitresults}
 \end{table}
 
The resummed version reads:
\begin{equation}
\label{eqn:resummedBeta}
\beta(Q)=-\frac{8}{5}abcQ^2\frac{\text{ sinh}(b(Q^2+1))\text{ sech}^{9/5}(b(Q^2+1))}{\left(1-a\text{ sech}^{4/5}(b(Q^2+1))\right)^2}.
\end{equation}
The predicted $\beta$-function, depicted in Fig.\ref{figure:beta}, has all the desired properties: it is negative, approaches zero at high and low energies, and has a minimum that characterizes the transition between the two regimes.

\section{Masses and decay constants}\label{Sec.MassesDecays}

The proposed dilaton Eq.(\ref{eqn:dilaton}) can be used to predict different meson Regge trajectory after solving their corresponding equation of motion \cite{Karch}. For vector mesons the five dimensional mass is equal to zero (in general it is given by $M_5^2=(\Delta-S)(\Delta+S-4)$, where $\Delta$ is the dimension of the operator dual to the field in QCD and $S$ is the spin). The corresponding equation of motion is:
\begin{equation}
\label{eqn:EOM}
\partial_z\left(\frac{e^{-\phi^{\Delta}(z)}}{z}\partial_z\psi_n(z)\right)=-M_n^2\frac{e^{-\phi^{\Delta}(z)}}{z}\psi_n(z).
\end{equation}
Here $M_n$ are interpreted as vector meson masses and the eigenfunctions $\psi_n(z)$ are normalized according to:
\begin{equation}
\int_0^{+\infty}dz\frac{e^{-\phi^{\Delta}(z)}}{z}|\psi_n(z)|^2=1.
\end{equation}
To solve the equation of motion Eq.(\ref{eqn:EOM}) the lightest mass must be fixed to be the lightest resonance in nature, the $\rho$-meson mass $M_0=M_{\rho}=775 \text{ MeV}$. To do so,  we impose $\lambda=0.4152$ in Eq.(\ref{eqn:dilaton}).

\begin{table}[h!]
\begin{tabularx}{0.48\textwidth} { 
  | >{\centering\arraybackslash}X 
  | >{\centering\arraybackslash}X 
  | >{\centering\arraybackslash}X
  | >{\centering\arraybackslash}X | }
 \hline
 $n$ & Scalar $M_n\text{ (MeV)}$ & Vector $M_n\text{ (MeV)}$ & Tensor $M_n\text{ (MeV)}$\\
 \hline
 0  & 944 & 775 & 1144\\
\hline
 1  & 1281 & 1154 & 1414\\
\hline
 2  & 1534 & 1427 & 1642\\
\hline
 3  & 1749 & 1652 & 1840\\
\hline
 4  & 1940 & 1850 & 2017\\
\hline
 5  & 2114 & 2028 & 2180\\
\hline
 6 & 2276 & 2192 & 2330\\
\hline
 7  & 2428 & 2344 & 2471\\
\hline
 8  & 2571 & 2487 & 2605\\
\hline
\end{tabularx}
\caption{Masses of scalar, vector and tensor mesons predicted by the DDSW model Eq.(\ref{eqn:dilaton}).}
\label{table:Masses}
\end{table}

\begin{figure}
\includegraphics[scale=0.55]{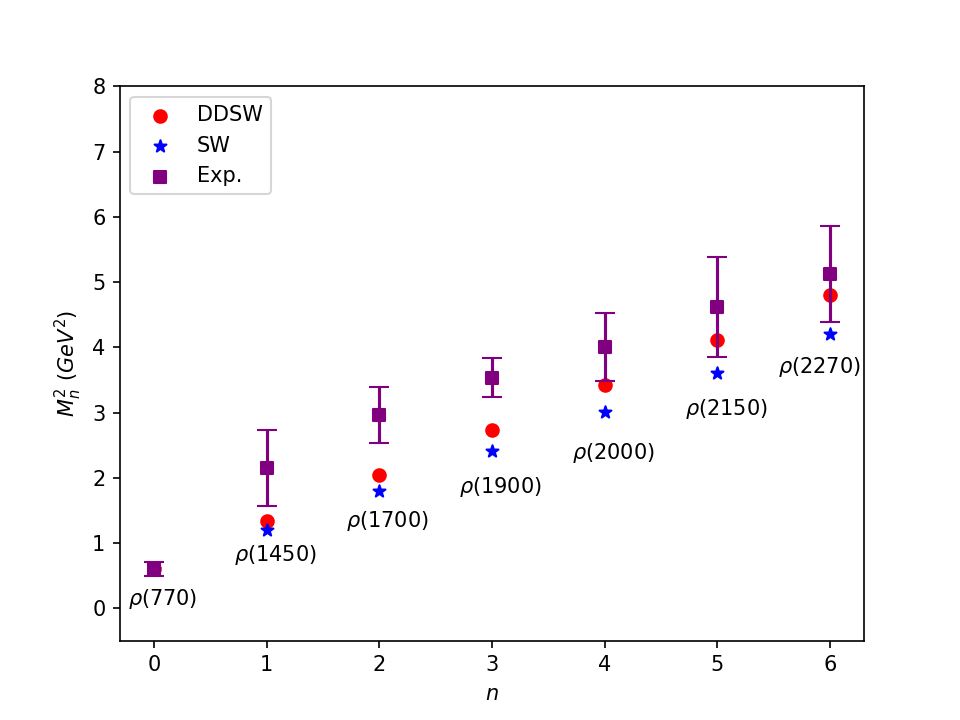}
\caption{Vector meson Regge trajectories $(M_n^2,n)$ for the DDSW model (red points), the SW model (blue stars), and comparison with experimental data (purple boxes) \cite{Workman}, \cite{Navas}.}
\label{figure:regge}
\end{figure}

Regge trajectories for our DDSW model are linear as in the Soft Wall model, that is, $M_n^2\sim n$.  
We depict our predictions using the DDSW model for the $\rho$-meson family (red points in Fig.\ref{figure:regge}) together with the physical values from the PDG \cite{Workman} (purple boxes) where errors are obtained via de half-width rule (i.e, $M_n^2 \pm M_n \Gamma_n$, with $\Gamma_n$ the total decay width of the particle \cite{Masjuan:2012sk}), and the prediction of Soft Wall model \cite{Karch} (blue stars).
Comparison to experimental data demands merging all $\rho$ states with total angular momentum $J=1$,  including 1S and 2S states, as we understand the holographic model cannot distinguish among them. In this respect, the Regge slope for the radial trajectory for the 7 known members of the $\rho$ family would read $0.87(7)$GeV$^{-2}$ as an average of the 1S and 2S trajectories from Ref.\cite{Masjuan:2012gc}. The DDSW model yields a slope $0.6962(18)$ GeV$^{-2}$.

The prediction of the $\rho$-meson decay constant using the DDSW model is also possible as is related to the residue at the pole $M_0$ on the equation of motion. It reads $F_{0}^{1/2}=F_{\rho}^{1/2}=346 \text{ MeV}$ which departs from the PDG value $F_{\rho}^{1/2}=348 \text{ MeV}$ by $0.57\%$ only \cite{Workman}. This is an improvement with respect to the Soft Wall (SW) result, which reads $F_{\rho}^{1/2}=260 \text{ MeV}$ \cite{Leutgeb}.
Other approaches such as the Semi-Hard Wall model (SHW, \cite{Kwee, Lebed}) or the Tachyon Condensation model (TC \cite{Casero, Iatrakis, Paredes}) have linear Regge trajectories only for large $n$. The departure for $F_{\rho}^{1/2}$ to the PDG one in these cases reaches $9.7\%$ and $10\%$, respectively (see Table \ref{table:Frho}).
\begin{table}[h!]
\begin{tabularx}{0.48\textwidth} { 
  | >{\centering\arraybackslash}X 
  | >{\centering\arraybackslash}X 
  | >{\centering\arraybackslash}X
  | >{\centering\arraybackslash}X | }
 \hline
 Model & $F_{\rho}^{1/2}\text{ (MeV)}$ & Departure\\
 \hline
 DDSW (this work)  & 346 & 0.57\% \\
\hline
 SW \cite{Karch}& 260 & 25\% \\
\hline
 SHW  \cite{Kwee, Lebed}  & 314 & 9.7\% \\
\hline
 TC \cite{Casero, Iatrakis, Paredes}& 313.2 & 10\%\\
\hline
\end{tabularx}
\caption{Comparison of $F_{\rho}^{1/2}$ in DDSW (this work) and other different holographic QCD models. We are comparing with the experimental value $F_{\rho}^{1/2}=348 \text{ MeV}$ \cite{Workman} and using  $\text{Departure}=|\text{prediction}-\text{experiment}|/\text{experiment}$.}\label{table:Frho}
\end{table}

From the $\rho$ decay constant predicted by our DDSW, we can use the relation
\begin{equation}
F_{\rho}^2=\frac{3M_{\rho}^3}{4\pi \alpha^2}\Gamma(\rho^0\to e^+e^-).
\end{equation}
to obtain $\Gamma(\rho^0\to e^+e^-)=6.87 \text{ keV}$. This gives a departure of $2.41\%$ in comparison to the experimental value $7.04(6)\text{ keV}$ \cite{Workman}. For the usual Soft Wall model the value is $2.19 \text{ keV}$ with a departure of $68.9\%$.

\begin{figure}
\includegraphics[scale=0.55]{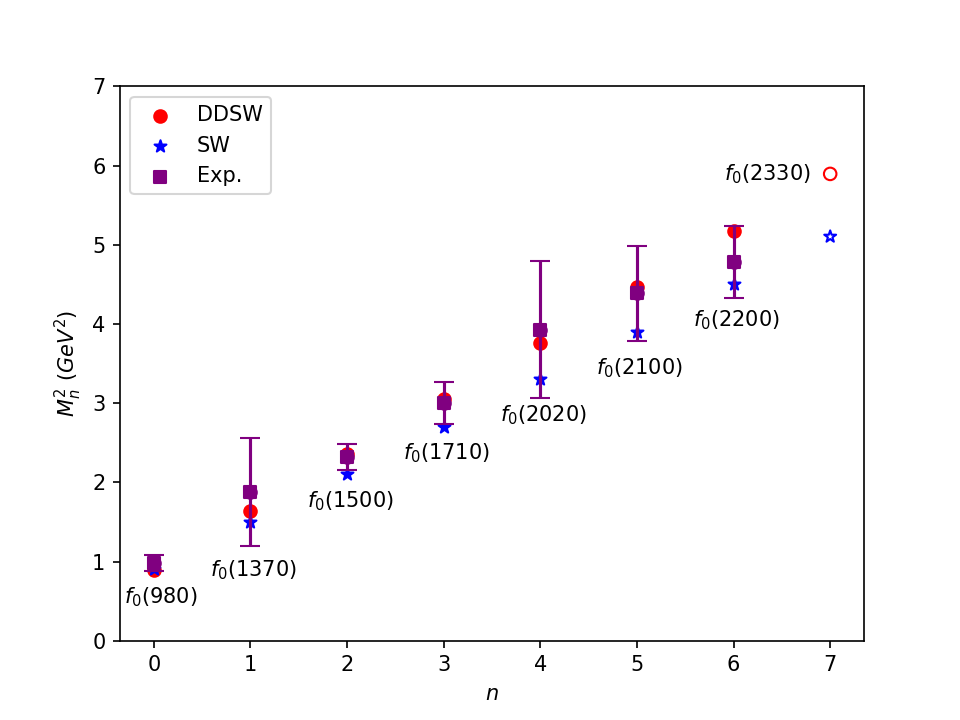}
\caption{Scalar meson Regge trajectories $(M_n^2,n)$ for the DDSW model (red points), the SW model (blue stars), and comparison with experimental data (purple boxes) \cite{Navas}.}
\label{figure:reggeScalar}
\end{figure}

Using the same procedure we can find the spectrum of scalar and tensor mesons. Scalar mesons are particularly interesting since they have a non zero five dimensional mass $M_5^2=-3$. Following the usual scalar field action \cite{Colangelo}, we must solve:
\begin{equation}
\label{eqn:EOMScalar}
\begin{aligned}
\partial_z\left(\frac{e^{-\phi^{\Delta}(z)}}{z^3} \right. & \left. \vphantom{\frac{e^{-\phi^{\Delta}(z)}}{z^3}} \partial_z\psi_n(z)\right) + \frac{3}{z^5}e^{-\phi^{\Delta}(z)}\psi_n(z) \\
&= -M_n^2\frac{e^{-\phi^{\Delta}(z)}}{z^3}\psi_n(z),
\end{aligned}
\end{equation}
with the normalization:
\begin{equation}
\label{eqn:scalarNormalization}
\int_0^{+\infty}dz\frac{e^{-\phi^{\Delta}(z)}}{z^3}|\psi_n(z)|^2=1.
\end{equation}
Results are depicted in the second column of Table \ref{table:Masses}, in which an improvement respect the original Soft Wall model is achieved, and more accurate results are obtained when comparing to experimental data \cite{Navas}. In fact,  resonances with $n=2, 3, 5$ show a departure to the experimental counterpart below $1\%$. In Figure \ref{figure:reggeScalar}, the empty circle and the empty star represent the prediction of DDSW and Soft Wall models, respectively, for $f_0(2330)$ meson. We have not included this resonance as a experimental data point since there is no consensus for its mean value \cite{Navas}. In this way we predict the mass of the $f_0(2330)$ as $2428 \text{ MeV}$ in the DDSW model. We observe this value is similar to that of \cite{Rodas}, which gives a mass of $2419\pm 64\text{ MeV}$, heavier than other analysis present in the PDG \cite{Navas}.

\begin{table}
\begin{tabularx}{0.48\textwidth} { 
  | >{\centering\arraybackslash}X 
  | >{\centering\arraybackslash}X 
  | >{\centering\arraybackslash}X
  | >{\centering\arraybackslash}X | }
 \hline
 $n$ & Scalar $F_n^{1/2}\text{ (MeV)}$ & Vector $F_n^{1/2}\text{ (MeV)}$ & Tensor $F_n^{1/2}\text{ (MeV)}$ \\
 \hline
 0  & 289 & 346 & 0.852\\
\hline
 1  & 317 & 393 & 106\\
\hline
 2  & 346 & 435 & 122\\
\hline
 3  & 371 & 469 & 137\\
\hline
 4  & 390 & 496 & 150\\
\hline
 5  & 408 & 519 & 163\\
\hline
 6 & 423 & 540 & 175\\
\hline
 7  & 438 & 558 & 186\\
\hline
 8  & 457 & 575 & 202\\
\hline
\end{tabularx}
\caption{Decay constants of scalar, vector and tensor mesons predicted by the DDSW model.}
\end{table}

\begin{figure}
\includegraphics[scale=0.55]{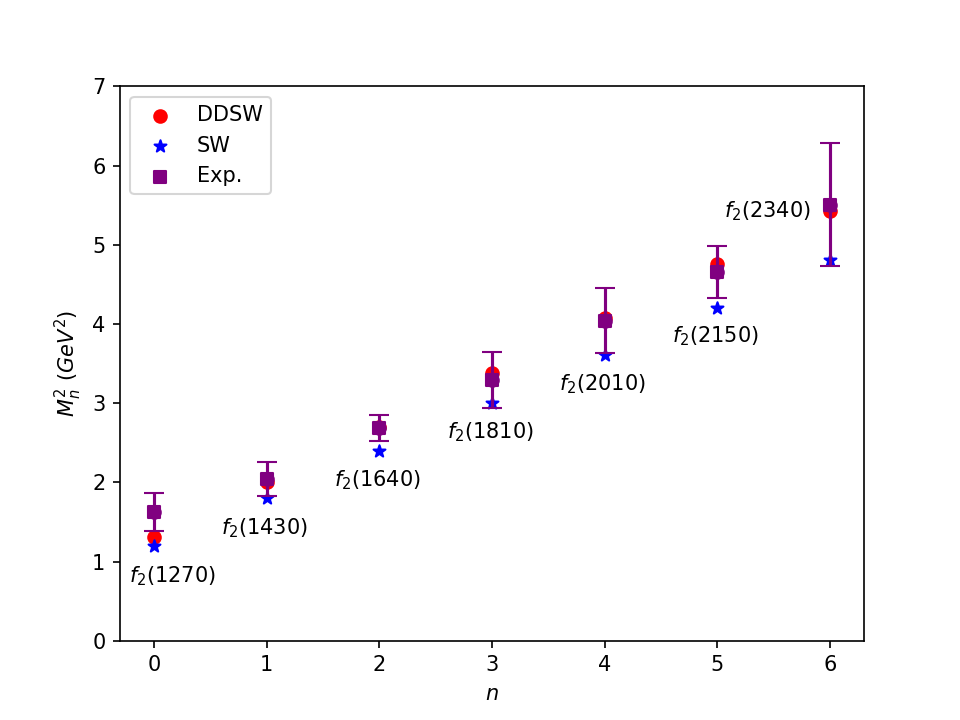}
\caption{Proposal for tensor meson Regge trajectory $(M_n^2,n)$ for the DDSW model (red points), the SW model (blue stars), and comparison with experimental data (purple boxes)\cite{Navas}.}
\label{figure:reggeTensor}
\end{figure}

Finally we adress the spectrum of tensor mesons. Here $M_5^2=0$, so the corresponding equation of motion is:
\begin{equation}
\label{eqn:EOMTensor}
\partial_z\left(\frac{e^{-\phi^{\Delta}(z)}}{z^3}\partial_z\psi_n(z)\right)=-M_n^2\frac{e^{-\phi^{\Delta}(z)}}{z^3}\psi_n(z),
\end{equation}
with the same normalization as in the scalar case Eq.(\ref{eqn:scalarNormalization}). Some studies of tensor mesons distinguish four different Regge trajectories \cite{MasjuanArriolaBroniowski}, \cite{Anisovich}. We have introduced tensor mesons with the same action as in SW model \cite{Karch}, \cite{Mamedov} introducing our dilaton field. The output of our model is a single trajectory, so we have compared our predicted masses with the experimental data available. More precisely, our model is a large-$N_c$ model, so we have stable mesons, with zero decay widths. As consequence, we have grouped the resonances found in experiment if they are overlaped by its decay width. For example, $f_2(1950)$ has a large decay width ($464\pm 24 \text{ MeV}$, \cite{Navas}) that overlaps with other resonances like $f_2(2010)$. In this case we have plotted only one resonance, the $f_2(2010)$, instead of two, by comparing with our predictions from DDSW model. In this way, the Regge trajectory in Table \ref{figure:reggeTensor} is a proposal of organising the $f_2$ resonances.

As a summary of results, we conclude that thanks to including not only the $z^2$ behavior at large $z$ -which renders the Regge trajectories linear- but also the $z\to 0$ constant behavior of DDSW dilaton, our mass predictions are closer to their experimental counterparts compared to the original SW model.

\section{Conclusions}\label{Sec.Conclusions}

In this work, we have explored a new holographic model of QCD, named Double Dilaton Soft Wall model (DDSW), which from the Ricci flow allow us to obtain a parameterization for $\alpha_s$ able to match an infrared fixed point at low energies with pQCD at high energies. We used an AdS/CFT dictionary formula defined in Eq.(\ref{eqn:AdSCFTdictionary}), to build the DDSW raising from a dilaton background defined in Eq.(\ref{eqn:dilaton}) that breaks \textbf{chiral} symmetry geometrically thanks to including two dilaton fields of opposite signs (cf. \cite{Nicotri,Zuo,Afonin}), each of them understood as vector and axial-vector mesons seeing different geometries \cite{Hirn}.
The matching with pQCD, which is found way above 2 GeV, is successful thanks to including in the holographic side large-$N_c$ corrections. In the process, we use experimental data from Ref. \cite{Deur} to fit the free leftover parameters of the $\alpha_s$ parameterization. Because our model Eq.(\ref{eqn:Resum}) have various adjustable parameters which limits its predictive power, we expect to improve it in the future.

As a by-product, we have studied the spectrum of vector mesons in the DDSW model by solving the equation of motion \cite{Karch} of different meson families with the dilaton mentioned above. This allowed us to determine masses as eigenvalues of the differential operator defined by the equation of motion, obtaining linear Regge trajectories for scalar, vector and tensor mesons. In particular, the predicted scalar and vector meson masses are heavier than in the usual Soft Wall model, thus closer to experimental results. Based on these results, we have done a proposal for the tensor meson Regge trajectory. Moreover, the squared root of the decay constant of the $\rho$-meson is improved compared to usual SW models, giving a departure of less than $1\%$ compared to the experimental result.

These results open the possibility of exploring other phenomenological results. In particular, our method could be applied to other dilaton backgrounds, obtaining strong coupling runnings at low energies and corresponding matching to pQCD.

\section*{Acknowledgements}
We are grateful to V. Vento for discussions and encouregement.

The work of PM has been supported by the Ministerio de Ciencia e Innovación under grant PID2020-112965GB-I00 and by the Secretaria d’Universitats i Recerca del Departament d’Empresa i Coneixement de la Generalitat de Catalunya under grant 2021 SGR 00649. The CERCA program of the Generalitat de Catalunya partially funds IFAE.

\nocite{*}
\bibliographystyle{unsrt}

\end{document}